\documentclass[aps,prb,twocolumn,amsmath,amssymb]{revtex4}
\usepackage{graphicx} 
\usepackage{bm}
\usepackage{dcolumn}
\usepackage{amsmath}
\usepackage{epstopdf}
\usepackage{natbib}

\begin{document}

\title{Excitonic gap formation in neutral bilayer structures}

\author{V. Apinyan}
\author{T. K. Kope\'{c}}
\email{v.apinyan@int.pan.wroc.pl}

\affiliation{%
Institute of Low Temperature and Structure Research (ILTSR), Polish Academy of Sciences, PO. Box 1410, 50-950 Wroclaw 2, Poland 
}%
\date{\today}

\begin{abstract}
%
We consider the pairing between conduction band electrons, and the valence band holes in the neutral bilayer-type structures. By employing the bilayer Hubbard model, we show the possibility of the inter-plane exciton formation in the system without applied external field. The in-plane and inter-plane Coulomb interaction effects on the pairing mechanism are considered, and the role of the in-plane particle hopping asymmetry on the gap behavior is analyzed in the paper. We show that both Frenkel-type pairing channel and Wannier-Mott-type excitonic pairings are present in the considered system. We analyze also the structure of the chemical potential in the bilayer system. The temperature effects, and the tunable inter-plane electron hopping effects are discussed. For the Frenkel channel, we have shown a particular behavior of the chemical potential at the very low temperatures, which is related to the degenerated Frenkel-gap.    
\end{abstract}
\maketitle

\section{\label{sec:Section_1} Introduction}
%
The physics related to bilayer structures is now the subject of intensive theoretical and experimental researches due to the recent developments in the field of electronic nanotechnologies and researches on the high speed information processing technologies. The most prominent example in the technological applications of such systems is related to the recently investigated graphene-based bilayers, which are very promising in the context of future nanoelectronic devices.\cite{cite-1,cite-2,cite-3,cite-4,cite-5,cite-6,cite-7,cite-8,cite-9} Another interest related to the double layer structured materials is that they are good candidates for the exciton condensate formation at the very low temperatures.\cite{cite-10,cite-11,cite-12,cite-13,cite-14,cite-15,cite-16,cite-17,cite-18,cite-19} The possibility of formation of the excitonic condensates in semiconducting materials has been suggested long years before \cite{cite-20,cite-21,cite-22,cite-23, cite-24} and, it was an intensively studied problem in the intervening years.\cite{cite-25,cite-26,cite-27,cite-28,cite-29,cite-30,cite-31}   

To control the formation of the excitonic condensate in the double layer structured systems, an external electric or magnetic field are supposed to be applied to the bilayer structure. The possibility of the excitonic condensate formation in two-dimensions (2D) , under strong magnetic field in a semiconductor heterostructure, with a very field-sensitive band structure, is given in Refs. \onlinecite{cite-32}  and \onlinecite{cite-33}. A thermodynamically stable condensate formation in InAs-GaSb -based system is discussed in Ref.\onlinecite{cite-12}, where a study of the Keldysh-BCS theory is given. Meanwhile, a new platform-material is proposed for studying the excitonic condensates.    
Recently, the detailed phase diagram for the inverted InAs/GaSb bilayer quantum wells is given also,\cite{cite-17} and the possibility of the stable $s$-wave exciton condensed phase is proved. Contrary, for the large interlayer tunneling amplitude, a topologically non-trivial quantum spin Hall insulator phase is obtained with the $p$-wave pairing gap. 

Despite of many theoretical investigations in this field of research the exact results, for the ground state properties of the electron-hole systems are rare. A determinant quantum Monte Carlo study of the exciton condensation in the bilayer Hubbard model is given recently. \cite{cite-18} Both, on-site Coulomb interaction and interlayer electrostatic interactions are considered.However, the question whether the true excitonic condensate formation is possible at the very low temperatures is still open.   

Another interesting problem in the field of the correlated bilayer systems \cite{cite-34} is related to the phase coherence in the excitonic condensate regime. Mainly, it has been shown recently that the excitonic pair formation and the excitonic condensation, are two different phases of matter. \cite{cite-30,cite-31,cite-35,cite-36} Thus, the question of the coherent excitonic condensate formation in double layer structures needs to be revised substantially. 

In this paper we examine the possibility of the exciton formation in the asymmetric double bilayer electron-hole structure with zero applied external electric field. We suppose the case of the negative hopping amplitude for the layer-1 electrons (see in  Fig.~\ref{fig:Fig_1}), rending the problem to that of the electron-hole type bilayer structure. We will show how the excitonic gap is formed and how the strength of it determines the type of the excitons, which are divided into two categories: Frenkel and Wannier-Mott-type excitons. Frenkel-type excitons are mediated by the localized Coulomb interaction and are the strongly bound excitations, and the Wannier-Mott excitons are delocalized, controlled by a weak Coulomb attraction between the electron-hole particles. For a asymmetric type bilayer structure we demonstrate that the solutions of the chemical potential  appear in the form of two well defined bands, separated by a gap region and we show that the values of energies, at which the Frenkel-type formations could appear in the system, are located between different bounds of the chemical potential forming the Frenkel channel in the exciton formation energy scheme. In the region of the very low pairing gap values, we show the presence of the Wannier-Mott-type excitonic gap lobes, which decrease when increasing the intralayer Coulomb interaction. 

We show, that at zero temperature the excitonic gap parameter is degenerated,  which is a direct consequence of the presence of the Frenkel-type energy zone in the chemical potential solution and which disappears at higher temperatures, when the single valued Frenkel channels are opening. We will calculate the temperature dependence of the excitonic gap parameter by considering different values of the intralayer Coulomb interaction parameter $U/|t_{1}|$.         

The paper is organized as follows: In the Section \ref{sec:Section_2} we introduce the bilayer Hamiltonian to describe the possible interactions in the system. We discuss also the linearization of the non-linear interaction terms in the Hamiltonian and the general solution strategy of the problem.
In the Section \ref{sec:Section_3} we introduce the effective fermionic action and we introduce the self-consistent (SC) equations for the excitonic gap parameter and chemical potential. In Section \ref{sec:Section_4}, we give the numerical solution of the problem of the interacting bilayer system, and we present the results for the excitonic gap parameter and the chemical potential. The temperature dependences of calculated quantities are also given there. Finally, in the Section \ref{sec:Section_5} we give the conclusion to obtained results.
 %
\section{\label{sec:Section_2} The method}
\begin{figure}
\begin{center}
\includegraphics[scale=.65]{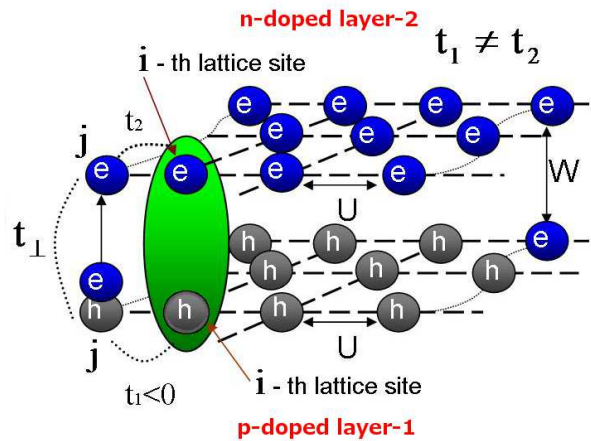}
\caption{\label{fig:Fig_1}(Color online) The electron-hole bilayer structure. p-doped hole layer (see the layer-1) and n-doped electron layer (see the layer-2) are represented. The outermost diagonal sites in the layer-1 represent two-processes: the electron hopping, from layer-1 to layer-2 (see at the left diagonal lattice site) and the electron-electron interlayer interaction, before hopping process (see at the right diagonal lattice site).}
\end{center}
\end{figure}

%
We introduce the Hamiltonian of our bilayer system with the in-plane and inter-plane couplings and we discuss the theoretical formalism. The bilayer is composed of two square lattices numbered as $l=1,2$, doped respectively with electrons and holes (as it is shown in Fig.~\ref{fig:Fig_1}). The Hamiltonian is 
\begin{eqnarray}
&&H=-\sum_{\substack{\left\langle i, j \right\rangle\\ l,\sigma}}t_{l}\left(c^{\dag}_{il\sigma}c_{jl\sigma}+h.c.\right)-t_{\perp}\sum_{i,\sigma}\left(c^{\dag}_{i1\sigma}c_{i2\sigma}+h.c.\right)
\nonumber\\
&&-\sum_{i,l,\sigma}\mu_{l}n_{il\sigma}+U\sum_{i,l}\left[(n_{il\uparrow}-1/2)\cdot(n_{il\downarrow}-1/2)-1/4\right]
\nonumber\\
&&+W\sum_{i,\sigma,\sigma'}\left[(n_{i1\sigma}-1/2)\cdot(n_{i2\sigma'}-1/2)-1/4\right].
\label{Equation_1}
\end{eqnarray}
Here, $\left\langle i, j \right\rangle$ denotes the sum over the nearest neighbors lattice sites, $t_{l}, l=1, 2$ is the in-plane hopping amplitude and $t_{\perp}$ is the inter-plane hopping amplitude. For a simple treatment, we consider the double layer structure without applied external electric field and we suppose that the chemical potentials in both layers are opposite in sign $\mu_{1}=-\mu_{2}$. This, in turn, will introduce the p-doping in one layer and n-doping into another. The on-site Hubbard-$U$ term in Eq.(\ref{Equation_1}) is written in the way that the case $\mu=0$ corresponds to the half-filling. The parameter $U$ is the local intralayer Coulomb interaction. We consider the double layer with respect to the half-filling conditions in each layer $\left\langle n_{l} \right\rangle=1$, for $l=1,2$. The formation of the excitons in the layered structure and the possibility of their further condensation at the low temperatures requires the attraction between the electrons and holes. In order to describe the interaction between the layers in the structure, we include also the Hubbard interaction term between different layers and this is described by the last term in Eq.(\ref{Equation_1}). The parameter $W$, in Eq.(\ref{Equation_1}), is the local interlayer electron-electron Coulomb repulsion. This term will be responsible for the excitonic band formation.    
%
\subsection{\label{sec:Section_2_1} Decoupling of interactions}
%
In order to decouple the quadratic terms appearing in the expression of the bilayer Hubbard Hamiltonian we will write the terms in the Eq.(\ref{Equation_1}) in more convenient form, by shifting the chemical potential in both layers and by adding to it the interaction constants. Namely, we introduce $\bar{\mu}_{l}$, $l=1,2$, as $\bar{\mu}_{l}=\mu_{l}+U/2+W$. The intralayer Coulomb interaction term could be also written in more symmetric form, taking into account the spin-rotational invariance, namely
\begin{eqnarray}
n_{il\uparrow}n_{il\downarrow}=\frac{n^{2}_{il}}{4}-S^{2}_{lz}
\label{Equation_2}
\end{eqnarray}
with $n_{il}=n_{il\uparrow}+n_{il\downarrow}$ and
\begin{eqnarray} 
S_{lz}=\frac{1}{2}\left(n_{il\uparrow}-n_{il\downarrow}\right).
\label{Equation_3}
\end{eqnarray} 

Next, we will pass to the Grassmann representation for the fermionic variables and we write the partition function of the system by employing the imaginary-time fermionic path-integral method. \cite{cite-37}
For this, we introduce the imaginary-time variables $\tau$ at each lattice site $i$. The time variables $\tau$ vary in the interval $(0,\beta)$, where $\beta=1/T$ with $T$ being the temperature. The time-dependent variables $c_{il}(\tau)$ ($\bar{c}_{il}(\tau)$) are satisfying the anti-periodic boundary conditions for fermions $c_{il}(\tau+\beta)=-c_{il}(\tau)$.\cite{cite-38} Then, the grand canonical partition function of the system is 
\begin{eqnarray}
Z=\int\left[Dc^{\dag}Dc\right]e^{-S\left[c^{\dag},c\right]},
\label{Equation_4}
\end{eqnarray}
where, the action in the exponential is expressed as
\begin{eqnarray}
S\left[c^{\dag},c\right]=S_{B}\left[c^{\dag},c\right]+\int^{\beta}_{0}d\tau H\left(\tau\right).
\label{Equation_5}
\end{eqnarray}
The first term, in Eq.(\ref{Equation_5}), is the fermionic Berry-term. It is given as
\begin{eqnarray}
S_{B}\left[c^{\dag},c\right]=\sum_{i,l}\int^{\beta}_{0}d\tau c^{\dag}_{il}(\tau)\frac{\partial}{\partial \tau}c_{il}(\tau).
\label{Equation_6}
\end{eqnarray}
Furthermore, we will combine the quadratic and linear total density terms in Eq.(\ref{Equation_1}) and we will decouple the obtained non-linear term by the scalar-field Hubbard-Stratanovich linearization procedure.\cite{cite-37} For the total electron density quadratic term, we have
\begin{eqnarray}
&&\exp\left[-U/4\sum_{i,l}\left(n_{il}-\frac{2\bar{\mu}_{l}}{U}\right)^{2}\right]
\nonumber\\
&&=\int\left[{DV}\right]\exp\left[\sum_{i,l}-\frac{V^{2}_{il}}{U}+iV_{il}\left(n_{il}-\frac{2\bar{\mu}_{l}}{U}\right)\right],
\nonumber\\
\label{Equation_7}
\end{eqnarray}
next, for the electron density difference quadratic term, we have 
\begin{eqnarray}
&&\exp\left[\frac{U}{4}\sum_{i,l}\left(n_{il\uparrow}-n_{il\downarrow}\right)^{2}\right]
\nonumber\\
=
&&\int\left[D\Delta_{c}\right]\exp\left[\sum_{i,l}-\frac{\Delta^{2}_{cil}}{U}+\Delta_{cil}\cdot\left(n_{il\uparrow}-n_{il\downarrow}\right)\right].
\nonumber\\
\label{Equation_8}
\end{eqnarray}
We have introduced in Eqs.(\ref{Equation_7}) and (\ref{Equation_8}) two real field variables $V$ and $\Delta_{c}$, which couple to the total density and density difference terms, figuring in the initial Hamiltonian. After those transformations, the partition function of the system takes the following form:
\begin{eqnarray}
Z=\int\left[Dc^{\dag}Dc\right]\int\left[DV\right]\int\left[D\Delta_{c}\right]e^{-S_{0}\left[c^{\dag},c,V,\Delta_{c}\right]},
\label{Equation_9}
\end{eqnarray}
where, the action in the exponential is the total fermionic action of our system. It is given by 
\begin{eqnarray}
&&S_{0}\left[c^{\dag},c,V,\Delta_{c}\right]=S_{B}\left[c^{\dag},c\right]
\nonumber\\
&&+\sum_{\substack{\left\langle i, j \right\rangle\\ l,\sigma}}\int^{\beta}_{0}d\tau t_{l}\left(c^{\dag}_{il\sigma}(\tau)c_{jl\sigma}(\tau)+h.c.\right)
\nonumber\\
&&+t_{\perp}\sum_{i,\sigma}\int^{\beta}_{0}d\tau\left(c^{\dag}_{i1\sigma}(\tau)c_{i2\sigma}(\tau)+h.c.\right)
\nonumber\\
&&+\sum_{i,l}\int^{\beta}_{0}d\tau \left[\frac{V^{2}_{il}(\tau)}{U}-iV_{il}(\tau)\cdot\left(n_{il}(\tau)-\frac{2\bar{\mu}_{l}}{U}\right)\right]
\nonumber\\
&&+\sum_{i,l}\int^{\beta}_{0}d\tau\left[\frac{\Delta^{2}_{cil}(\tau)}{U}-\Delta_{cil}(\tau)\cdot\left(n_{il\uparrow}(\tau)-n_{il\downarrow}(\tau)\right)\right]
\nonumber\\
&&-W\sum_{i,\sigma,\sigma'}\int^{\beta}_{0}d\tau n_{i1\sigma}(\tau)\cdot n_{i2\sigma'}(\tau).
\label{Equation_10}
\end{eqnarray}

For the integration over the $V_{il}$-variables, we have the following scalar field integral
\begin{eqnarray}
\int\left[DV\right]e^{-\sum_{i,l}\int^{\beta}_{0}d\tau \left[\frac{V^{2}_{il}}{U}-iV_{il}\cdot\left(n_{il}(\tau)-\frac{2\bar{\mu}_{l}}{U}\right)\right]}
\nonumber\\
=const\cdot e^{-S\left[V^{S}\right]},
\nonumber\\
\label{Equation_11}
\end{eqnarray}
where the action in the exponential in the right hand side in Eq.(\ref{Equation_11}) is given after the saddle point evaluations for $V_{il}$. It is not difficult to see that
\begin{eqnarray}
V^{S}_{l}=\frac{iU}{2}\left(\bar{n}_{l}-\frac{2\bar{\mu}_{l}}{U}\right)
\label{Equation_12}
\end{eqnarray}
with $\bar{n}_{l}=\left\langle n_{l} \right\rangle$, $l=1,2$, being the quantum statistical fermion averages for the total particle numbers in each layer. Then the action $S\left[V^{S}\right]$ in Eq.(\ref{Equation_11}) has the form
\begin{eqnarray}
S\left[V^{S}\right]=-\frac{U}{2}\sum_{i,l}\int^{\beta}_{0}d\tau \left(\bar{n}_{il}-\frac{2\bar{\mu}_{l}}{U}\right)\times
\nonumber\\
\times\left({n}_{il}\left(\tau\right)-\frac{2\bar{\mu}_{l}}{U}\right).
\label{Equation_13}
\end{eqnarray}
The integration over the scalar field $\Delta_{cil}$ could be also done by the saddle point method. For the saddle point value of $\Delta_{cil}$ we get
\begin{eqnarray}
\Delta^{S}_{cl}=\frac{U}{2}\left\langle S_{ilz}(\tau)\right\rangle
\label{Equation_14}
\end{eqnarray}
and the corresponding action becomes 
\begin{eqnarray}
S\left[\Delta^{S}_{c}\right]=-\sum_{i,l,\sigma}\int^{\beta}_{0}d\tau (-1)^{\sigma}\Delta^{S}_{cl}n_{il\sigma}.
\label{Equation_15}
\end{eqnarray}
For the half-filling case, we will fix $\Delta^{S}_{c1}=\Delta^{S}_{c2}=U/2$. 
For further simplifications, we will write the interlayer interaction term in Eq.(\ref{Equation_10}) in a compact form, by introducing the following complex variables
\begin{eqnarray}
\Delta_{i,\sigma\sigma'}(\tau)=c^{\dag}_{2i\sigma}(\tau)c_{1i\sigma'}(\tau)
\label{Equation_16}
\end{eqnarray}
with all possible spin directions and their complex conjugates. The variables $\Delta_{i,\sigma\sigma'}(\tau)$ are linear in electron density. Then, the interlayer Coulomb interaction term will reads as 
\begin{eqnarray}
S\left[\Delta^{\dag},\Delta\right]=-W\sum_{i,\sigma,\sigma'}\int^{\beta}_{0}d\tau|\Delta_{i,\sigma\sigma'}(\tau)|^{2}.
\label{Equation_17}
\end{eqnarray}
This form of the interaction action is more convenient for further decoupling process. 
%
\subsection{\label{sec:Section_2_2} The interlayer interaction}
%
We can decouple the non-linear term in Eq.(\ref{Equation_17}) by introducing the complex Hubbard-Stratanovich decoupling fields $Q^{\dag}_{i,\sigma\sigma'}(\tau)$ and $Q_{i,\sigma\sigma'}(\tau)$ coupled to the complex linear variables $\Delta_{i,\sigma\sigma'}(\tau)$ and $\Delta^{\dag}_{i,\sigma\sigma'}(\tau)$. The transformation is 
\begin{eqnarray}
&&\exp\left[W\sum_{i,\sigma,\sigma'}\int^{\beta}_{0}d\tau|\Delta_{i,\sigma\sigma'}|^{2}\right]
\nonumber\\
&&=\int\left[DQ^{\dag}DQ\right]\exp\left[-\sum_{i,\sigma,\sigma'}\int^{\beta}_{0}d\tau\frac{1}{W}|Q_{i,\sigma\sigma'}(\tau)|^{2}\right.
\nonumber\\
&&\left.+Q^{\dag}_{i,\sigma\sigma'}(\tau)\Delta_{i,\sigma\sigma'}(\tau)+\Delta^{\dag}_{i,\sigma\sigma'}(\tau)Q_{i,\sigma\sigma'}(\tau)\right].
\nonumber\\
\label{Equation_18}
\end{eqnarray}
Furthermore, we will see that the decoupling fields $Q^{\dag}_{i,\sigma\sigma'}(\tau)$ and $Q_{i,\sigma\sigma'}(\tau)$ play the role of excitonic pairing gap parameters for different spin orientations. 
The action related to the interlayer interaction term will reads as
\begin{eqnarray}
&&S\left[\Delta^{\dag},\Delta,Q^{\dag},Q\right]=\sum_{i,\sigma,\sigma'}\int^{\beta}_{0}d\tau\left[\frac{|Q_{i,\sigma\sigma'}(\tau)|^{2}}{W}\right.
\nonumber\\
&&\left.+Q^{\dag}_{i,\sigma\sigma'}(\tau)\Delta_{i,\sigma\sigma'}(\tau)+\Delta^{\dag}_{i,\sigma\sigma'}(\tau)Q_{i,\sigma\sigma'}(\tau)\right].
\label{Equation_19}
\end{eqnarray}
Thus, the last term in Eq.(\ref{Equation_10}) will transforms as $S_{W}\left[c^{\dag},c\right]\rightarrow S\left[\Delta^{\dag},\Delta,Q^{\dag},Q\right]$. 
Next, we will write the partition function of the system
\begin{eqnarray}
Z=\int\left[Dc^{\dag}Dc\right]\int\left[DQ^{\dag}DQ\right]e^{-S\left[c^{\dag},c,Q^{\dag}, Q\right]},
\nonumber\\
\label{Equation_20}
\end{eqnarray}
where the action contains already the terms given in Eq.(\ref{Equation_19}):
\begin{eqnarray}
&&S\left[c^{\dag},c,Q^{\dag}, Q\right]=S_{B}[c^{\dag},c]+
\nonumber\\
&&+\sum_{\substack{\left\langle i, j \right\rangle\\ l,\sigma}}\int^{\beta}_{0}d\tau t_{l}\left(c^{\dag}_{il\sigma}(\tau)c_{jl\sigma}(\tau)+h.c.\right)
\nonumber\\
&&+t_{\perp}\sum_{i,\sigma}\int^{\beta}_{0}d\tau\left(c^{\dag}_{i1\sigma}(\tau)c_{i2\sigma}(\tau)+h.c.\right)
\nonumber\\
&&+S\left[V^{S}\right]+S\left[\Delta^{S}_{c}\right]+S\left[\Delta^{\dag},\Delta,Q^{\dag},Q\right].
\nonumber\\
\label{Equation_21}
\end{eqnarray}
In the next Section, we will use the form of the action given in Eq.(\ref{Equation_21}) to derive the expression for the excitonic gap parameter.
%
\section{\label{sec:Section_3} Effective action for fermions}
%
Now, it is easy to write the total action of the system given in Eq.(\ref{Equation_21}) in the Fourier transformed form, using the expression 
of the decoupled action in the Eq.(\ref{Equation_19}). To this end, we will introduce here the following four component Nambu spinor $\Psi_{{\bf{k}}}(\nu_{n})$
\begin{eqnarray}
\Psi_{{\bf{k}}}(\nu_{n})=\left[
\begin{array}{cccc}
&c_{{\bf{k}}_{1}\uparrow}(\nu_{n}) \\& c^{\dag}_{{\bf{k}}_{1}\downarrow}(\nu_{n}) \\& c_{{\bf{k}}_{2}\downarrow}(\nu_{n})  \\& c^{\dag}_{{\bf{k}}_{2}\downarrow}(\nu_{n})
\end{array}
\right]
\label{Equation_22}
\end{eqnarray}
and the complex conjugate Nambu vector $\Psi^{\dag}_{{\bf{k}}}(\nu_{n})$. Here ${\bf{k}}$, is the electron wave vector in the reciprocal Fourier space, and the fermionic Matsubara frequencies\cite{cite-38} $\nu_{n}={(2n+1)\pi/\beta}$, with $n=0,\pm 1,\pm2,...$.   
Then, we have 
\begin{eqnarray}
Z=\int\left[D\Psi^{\dag}D\Psi\right]\int\left[DQ^{\dag}DQ\right]e^{-S\left[\Psi^{\dag},\Psi,Q^{\dag},Q\right]},
\nonumber\\
\nonumber\\
\label{Equation_23}
\end{eqnarray}
where the action $S\left[\Psi^{\dag},\Psi,Q^{\dag},Q\right]$ in the exponential is 
\begin{eqnarray}
S\left[\Psi^{\dag},\Psi,Q^{\dag},Q\right]=\frac{1}{\beta{N}}\sum_{{\bf{k}},\nu_{n}}\Psi^{\dag}_{{\bf{k}}}(\nu_{n})\hat{G}^{-1}_{{\bf{k}}}\left(\nu_{n}\right)\Psi_{{\bf{k}}}(\nu_{n}).
\nonumber\\
\label{Equation_24}
\end{eqnarray}
Here, the matrix $\hat{G}^{-1}_{{\bf{k}}}\left(\nu_{n}\right)$ is a 4 $\times$ 4 matrix, given as 
\begin{eqnarray}
&&\hat{G}^{-1}_{{\bf{k}}}\left(\nu_{n}\right)=
\nonumber\\
&&=\left(
\begin{array}{ccccrrrr}
E_{1{\bf{k}}}\left(\nu_n\right) & 0 & {t}_{\perp}+Q_{\uparrow\uparrow} & 0\\
0 &-E_{1{\bf{k}}}\left(\nu_n\right)  & 0 & -{t}_{\perp}-Q^{\dag}_{\downarrow\downarrow} \\
{t}_{\perp}+Q^{\dag}_{\uparrow\uparrow} & 0 & E_{2{\bf{k}}}\left(\nu_n\right) & 0 \\
0 & -{t}_{\perp}-Q_{\downarrow\downarrow} & 0 & -E_{2{\bf{k}}}\left(\nu_n\right)  
\end{array}
\right).
\nonumber\\
\label{Equation_25}
\end{eqnarray}
The energy parameters $E_{l{\bf{k}}}\left(\nu_n\right)$, $l=1,2$, entering into the Eq.(\ref{Equation_25}), are given by
\begin{eqnarray}
E_{l{\bf{k}}}\left(\nu_n\right)=-i\nu_{n}-\Delta^{S}_{c}+2\tilde{t}_{l{\bf{k}}}+\bar{\mu}_{l}-\frac{U\bar{n}_{l}}{2}.
\label{Equation_26}
\end{eqnarray}
Next, $\tilde{t}_{l{\bf{k}}}$, $l$=1,2, are the renormalized in-plane hopping amplitudes $\tilde{t}_{l{\bf{k}}}=2t_{l}\gamma_{\bf{k}}$, and the energy dispersion is 
\begin{eqnarray}
\gamma_{\bf{k}}=\cos(k_{x})+\cos(k_{y}).
\label{Equation_27}
\end{eqnarray}
It is not difficult to derive the saddle point expression for the decoupling fields $Q_{i,\sigma\sigma'}$. Indeed, from Eqs.(\ref{Equation_19}) and (\ref{Equation_20}), we find that
\begin{eqnarray}
Q_{\sigma\sigma'}=-W\left\langle c^{\dag}_{2i\sigma}\left(\tau\right)c_{1i\sigma'}\left(\tau\right)\right\rangle,
\label{Equation_28}
\end{eqnarray}
and 
\begin{eqnarray}
Q^{\dag}_{\sigma\sigma'}=-W\left\langle c^{\dag}_{1i\sigma}\left(\tau\right)c_{2i\sigma'}\left(\tau\right)\right\rangle.
\label{Equation_29}
\end{eqnarray}
The quantum statistical averages in Eqs.(\ref{Equation_28}) and (\ref{Equation_29}) are given with the help of the fermionic action in Eq.(\ref{Equation_24}), mainly
\begin{eqnarray}
\left\langle ... \right\rangle=Z^{-1}\int\left[D\Psi^{\dag}D\Psi\right]\int\left[DQ^{\dag}DQ\right]...e^{-S\left[\Psi^{\dag},\Psi,Q^{\dag},Q\right]}.
\nonumber\\
\label{Equation_30}
\end{eqnarray}
In fact, the parameter $Q_{\sigma\sigma'}$ in Eq.(\ref{Equation_28}) plays the role of the excitonic gap, which is formed between the $p$-doped hole layer-1 and $n$-doped electron layer-2 (see in Fig.~\ref{fig:Fig_1}). Particularly, the gap $Q_{\uparrow\uparrow}$ describes the pairing gap between the $c^{\dag}_{2{\bf{k}}\uparrow}$ electron with the spin $\uparrow$ in the layer-2 and the $c_{1{\bf{k}}\uparrow}=d^{\dag}_{1{\bf{k}}\downarrow}$ hole with the spin $\downarrow$ in the layer-1 (we denoted by $d^{\dag}_{1{\bf{k}}\downarrow}$ the hole creation Grassmann operator). This corresponds to the conventional excitonic pairing picture as in the usual bulk semiconductors. \cite{cite-25,cite-26,cite-27,cite-28,cite-29,cite-35,cite-36} Contrary, the gap $Q_{\uparrow\downarrow}$ describes the pairing between the $c^{\dag}_{2{\bf{k}}\uparrow}$ electron with spin $\uparrow$ and the $c_{1{\bf{k}}\downarrow}=d^{\dag}_{1{\bf{k}}\uparrow}$ hole with the spin $\uparrow$. This is the result of the excitonic pairing with the possibility of the spin-reverse, when the electron is created in the layer-2. This type of pairing mechanism is rare, because of the fast e-h recombination effects that are always present in the considered system. For simplicity, we will not consider this type of pairing mechanism for the first treatment.

Without any loss of generality, we suppose the case of the pairing states with the uniform real gap parameters.\cite{cite-35,cite-36} Remember, that the interlayer pairing fields $Q^{\dag}$ and $Q$ are local in space and the part of the action corresponding them is then  
\begin{eqnarray}
S_{p}\left[Q^{\dag},Q,c^{\dag},c\right]=-\frac{1}{\beta{N}}\sum_{{\bf{k}},\nu_{n}}\sum_{\substack{\sigma,\sigma' \\ l,l'}}Q_{\sigma\sigma'}\times
\nonumber\\
\times \left(1-\delta_{ll'}\right)c^{\dag}_{{\bf{k}}l\sigma}c_{{\bf{k}}l'\sigma'}.
\nonumber\\
\label{Equation_31}
\end{eqnarray}
We have 
\begin{eqnarray}
Q_{\sigma\sigma'}=Q^{\dag}_{\sigma\sigma'}=-W\left\langle \Delta_{\sigma\sigma'}\right\rangle
\nonumber\\
=-\frac{W}{(\beta{N})^{2}}\sum_{{\bf{k}},\nu_{n}}\left\langle c^{\dag}_{2{\bf{k}}\sigma}(\nu_{n})c_{1{\bf{k}}\sigma'}(\nu_{n})\right\rangle.
\label{Equation_32}
\end{eqnarray}
We will rename here the gap parameters, as $Q_{\sigma\sigma'}\equiv \Delta_{\sigma\sigma'}$.
From the structure of the inverse Green function matrix given in Eq.(\ref{Equation_25}), it follows that $\Delta_{\sigma\sigma'}=\Delta_{\sigma\sigma}\delta_{\sigma\sigma'}$. We should mention here that this result is true only for the bilayer without applied external electric field, such as in our consideration. For the general case, when an external electric field is present, then $\Delta^{\rm ex}_{\uparrow\uparrow}\neq\Delta^{\rm ex}_{\downarrow\downarrow}$ and $|\mu_{1}|\neq|\mu_{2}|$. We will not consider here this case and it will be the subject for our future considerations. After calculating average in Eq.(\ref{Equation_32}) and taking into account the half-filling condition for the total particle densities in each layer, we have found the following system of coupled SC, non-linear equations
\begin{eqnarray}
&&\frac{1}{N}\sum_{{\bf{k}}}n_{F}\left(\epsilon^{+}_{{\bf{k}}}\right)+n_{F}\left(\epsilon^{-}_{{\bf{k}}}\right)=1,
\label{Equation_33}
\newline\\
&&\Delta=-\frac{W}{2N}\sum_{{\bf{k}}}\frac{\left(\tilde{t}_{\perp}+\Delta\right)\cdot\left[n_{F}\left(\epsilon^{+}_{{\bf{k}}}\right)+n_{F}\left(\epsilon^{-}_{{\bf{k}}}\right)\right]}{\sqrt{F^{2}_{{\bf{k}}}+\left({t}_{\perp}+\Delta\right)^{2}}},
\nonumber\\
\label{Equation_34}
\end{eqnarray}
where $n_{F}\left(z\right)$ is the Fermi-Dirac distribution function $n_{F}\left(z\right)=1/(e^{\beta{z}}+1)$ and the parameter $F_{{\bf{k}}}$ is
\begin{eqnarray}
F_{{\bf{k}}}=\tilde{t}_{1{\bf{k}}}-\tilde{t}_{2{\bf{k}}}-\mu.
\label{Equation_35}
\end{eqnarray}
For the convenience, we have omitted the spin indexes near the excitonic gap parameter $\Delta$.
Furthermore, the parameters $\epsilon^{\pm}_{{\bf{k}}}$, are given by the following relations
\begin{eqnarray}
\epsilon^{\pm}_{{\bf{k}}}=-\frac{U}{2}+2\left(t_{1}+t_{2}\right)\epsilon_{\bf{k}}+W\pm\sqrt{F^{2}_{{\bf{k}}}+\left({t}_{\perp}+\Delta\right)^{2}}.
\nonumber\\
\label{Equation_36}
\end{eqnarray}
The chemical potential $\mu$ is determined so as to maintain the average particle density at half-filling
and this is represented by the first equation Eq.(\ref{Equation_33}) in the system of SC equations.  

\section{\label{sec:Section_4} The results}
%
\subsection{\label{sec:Section_4_1} Gap solution}
\begin{figure}
\begin{center}
\includegraphics[scale=.5]{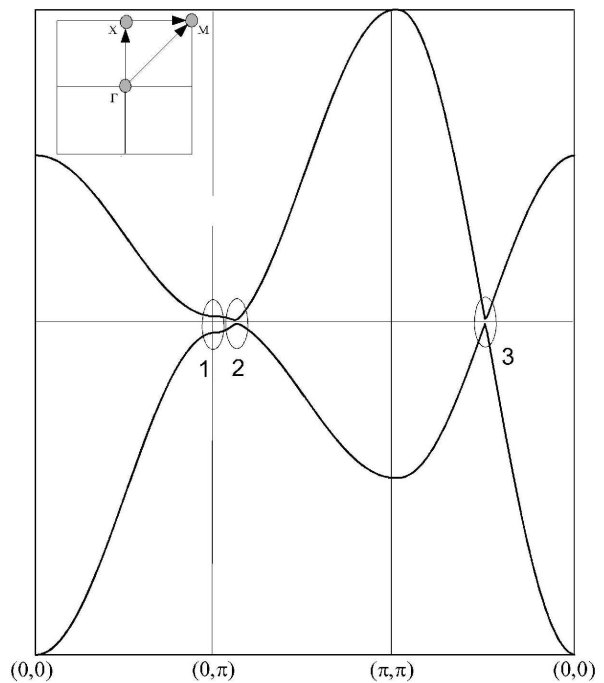}
\caption{\label{fig:Fig_2}(Color online) The interacting band structure, and charge gap formation in the bilayer system (see the Eq.(\ref{Equation_36})). The inset shows the symmetry points on a 2D square lattice.}
\end{center}
\end{figure}

\begin{figure}
\begin{center}
\includegraphics[width=250px,height=400px]{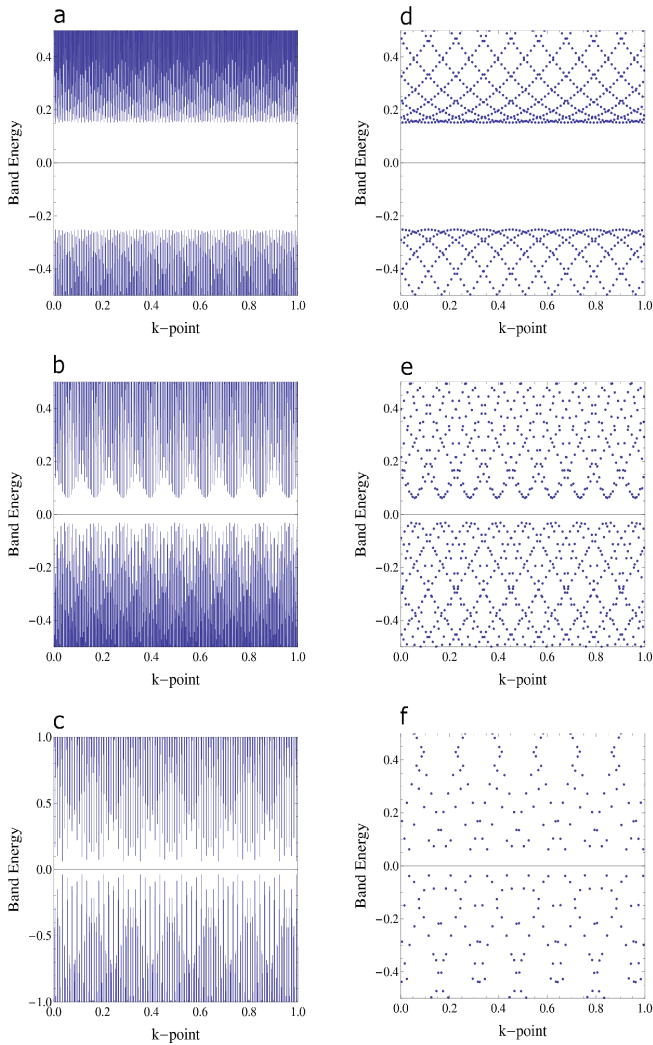}
\caption{\label{fig:Fig_3}(Color online) The Monkhorst-Pack grids for the interacting band structure along the high symmetry directions: $\Gamma \rightarrow X$, $X \rightarrow M$, $M \rightarrow \Gamma$ (from left to right). The upper panels the grid structure is shown with connected points, and in the lower panels the same structure is shown, with unconnected grid-points. }
\end{center}
\end{figure}

%
In this Section we present the numerical solution of the system of coupled SC equations given in Eqs.(\ref{Equation_33}) and (\ref{Equation_34}). We will perform the solution of this system of equations by employing the finite-difference approximation method in which the fast-convergent Newton's method \cite{cite-39} is employed for the nonlinear equations. The accuracy of convergence for numerical solutions is achieved with a relative error of order of $10^{-7}$. The ${\bf{k}}$-summations in Eqs.(\ref{Equation_33}) and (\ref{Equation_34}) were performed with the $(100\times100)$ $k$-points in the First Brillouin Zone (FBZ) on a simple square lattice layers. In Fig.~\ref{fig:Fig_2} we have presented the band energy dispersion curves for the interacting bilayer system. The calculation was performed for the fixed value of the intralayer Coulomb interaction $U/|t_{1}|=1.6$. The interlayer Coulomb interaction is chosen $W/|t_{1}|=0.75$ with the interlayer hopping amplitude $t_{\perp}=0.04|t_{1}|$ and we have sampled the FBZ by a number of 100 $k$-points. The functions in Eq.(\ref{Equation_36}) are represented on a discrete equally-spaced mesh of the sample square lattice. We see in In Fig.~\ref{fig:Fig_2}, that very small charge gaps $\Delta G =\epsilon_{1{\bf{k}}}-\epsilon_{2{\bf{k}}}$ are opening in the system along the high symmetry directions: $\left(0,0\right)\rightarrow \left(0,\pi \right)$ (at the boundary, the region numbered by 1),   $\left(0,\pi\right)\rightarrow \left(\pi,\pi \right)$ (see the region numbered by 2) and $\left(\pi,\pi\right)\rightarrow \left(0,0 \right)$ (see the region numbered by 3). The inset, in Fig.~\ref{fig:Fig_2}, shows the high symmetry points on a 2D square lattice. In Fig.~\ref{fig:Fig_3} we have shown the Monkhorst-Pack grids for band structure presented in Fig.~\ref{fig:Fig_2}. The left panels in the picture represent the Monkhorst-Pack scheme, when the band structure grid-points are connected. The panels from top to bottom represent the Monkhorst-Pack sets when evaluating from: $\gamma \rightarrow X$ (the left-top a panel in Fig.~\ref{fig:Fig_2}), $X \rightarrow M$ (the left-middle panel in Fig.~\ref{fig:Fig_2}) and $M \rightarrow \Gamma$ (the left-bottom panel in Fig.~\ref{fig:Fig_2}). The right panels in Fig.~\ref{fig:Fig_2} show the Monkhorst-Pack sets, when the grid-points are unconnected. The same high symmetry directions are considered when evaluating from top-right to bottom-right panels. We see clearly in all panels in Fig.~\ref{fig:Fig_2} (especially it is visible in the panels with the connected grid-points) that a charge-gap like small gap is always present in the band structure and it is diminishing, when evaluating along high symmetry directions clockwise (i.e. $\Gamma \rightarrow X \rightarrow M \rightarrow \Gamma$).    

\begin{figure}[htb]
\begin{center}
 \includegraphics[scale=.55]{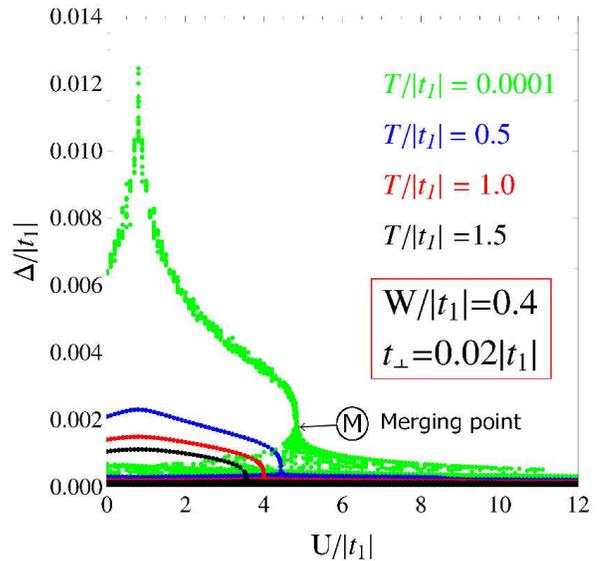}
 \caption{\label{fig:Fig_4} Excitonic gap parameter normalized to the electron layer hopping amplitude, as a function of the in-plane Coulomb interaction parameter $U/|t_{1}|$. Different values of the temperature are considered.}
\end{center}
\end{figure}

\begin{figure}[htb]
\begin{center}
 \includegraphics[scale=.55]{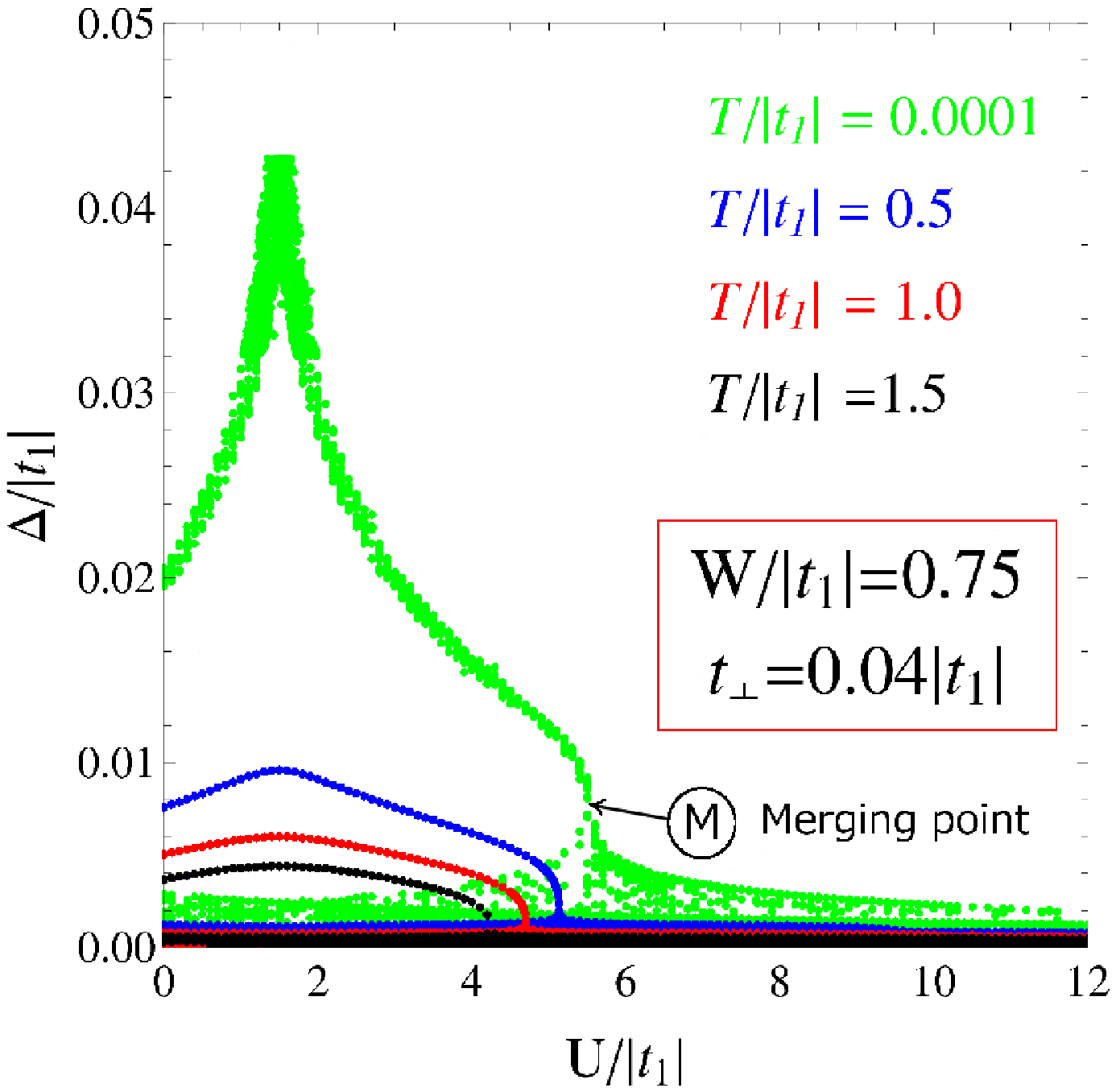}
 \caption{\label{fig:Fig_5} Excitonic gap parameter normalized to the electron layer hopping amplitude, as a function of the in-plane Coulomb interaction parameter $U/|t_{1}|$. Different values of the temperature are considered.}
\end{center}
\end{figure}

In Fig.~\ref{fig:Fig_4} we have presented the $U$-dependence of the excitonic gap parameter $\Delta$ for different values of the hopping-normalized temperature $T/|t_{1}|$. For the energy parameters in Fig.~\ref{fig:Fig_4}, we have considered the following values $W/|t_{1}|=0.4$, $t_{1}=-1.0$, $t_{2}=0.5|t_{1}|$ and $t_{\perp}=0.02|t_{1}|$. The $|t_{1}|$ energy-parameter serves as the unit of energy in our calculations. We have solved the system of equations in Eqs.(\ref{Equation_33}) and (\ref{Equation_34}) for the case of the electron hopping-asymmetry in different layers of the system.
%
%
%
Namely, we suppose the negative value for the lower layer-1 hopping parameter $t_{1}$, associating by this the layer-1 with a layer of hole particles. The interlayer hopping parameter $t_{\perp}$ is chosen to be small in order to not suppress significantly the exciton gap formation. It is clear in Fig.~\ref{fig:Fig_4}, that the excitonic gap is present in the case of the absence of the applied external field to the system. Thus, $\Delta$ is a spontaneous local gap formed between the layers in the system and we see in Fig.~\ref{fig:Fig_4} that it is non zero, practically for all values of the in-plane Coulomb interaction parameter $U/|t_{1}|$. The existence of the finite gap parameter, at $U/|t_{1}|=0$, is the fact of the dominant role of the interlayer Coulomb interaction $W$. For the intermediate values of the intralayer interaction $U/|t_{1}|$, the gap function endeavor to its maximal values. We observe also a specific gap-``see'' formation at the lower parts of the plots. Indeed, the upper solution lines in Fig.~\ref{fig:Fig_4} represent the strong binding Frenkel-type excitons, while the lowest energy solutions, in the form of a gap-``see'', represent the all possible Wannier-Mott-type excitons in the system. Thus, both type of excitons are possible in our bilayer structure in the case of the zero applied field to the structure. In Fig.~\ref{fig:Fig_4} the gap functions are plotted for the higher value of the interlayer Coulomb interaction parameter $W/|t_{1}|=0.75$ and also for the higher inter-plane hopping amplitude $t_{\perp}=0.02|t_{1}|$. We see in Figs.~\ref{fig:Fig_4} and ~\ref{fig:Fig_5}, that there is a smooth passage from Frenkel-gap to the Wannier-Mott limit (for example at the value $U/|t_{1}|\sim 5.5$, for $T/|t_{1}|=0.0001$ and for $W/|t_{1}|\sim 0.75$). The merging points ``M'', indicated in the circles in Figs.~\ref{fig:Fig_4} and ~\ref{fig:Fig_5}, correspond to the points, where the two type of solutions are joined.  

As we will see later on, in this Section, the upper solutions for the gap in Figs.~\ref{fig:Fig_4} and ~\ref{fig:Fig_5} correspond to the relatively small values of the chemical potential, while the Wannier-Mott gap region is related to the high values of the chemical potential. In the small and medium intralayer interaction regions, both type of excitons coexist. Contrary, for strong intralayer Coulomb interaction values, the Frenkel-gap disappears and the fluctuating Mott-gap stills open until very high values of the interaction parameter $U/|t_{1}|$. 

\begin{figure}[htb]
\begin{center}
 \includegraphics[scale=.35]{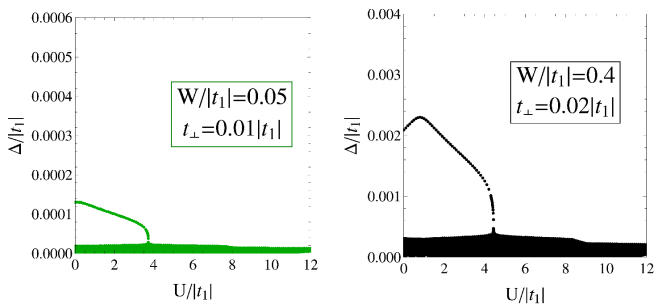}
 \caption{\label{fig:Fig_6} Excitonic gap parameter for small (left panel) and intermediate (right panel) values of the interlayer Coulomb interaction parameter $W/|t_{1}|$. The temperature is fixed at $T/|t_{1}|=0.5$.}
\end{center}
\end{figure}

\begin{figure}[htb]
\begin{center}
 \includegraphics[scale=.35]{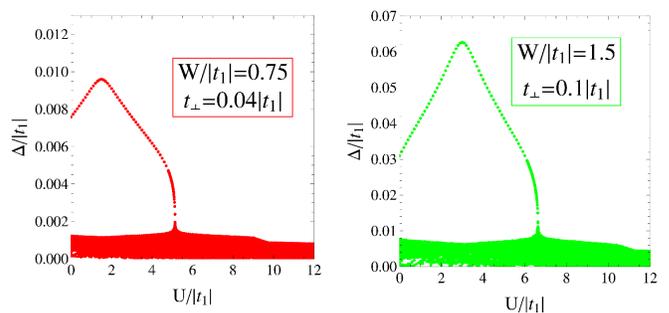}
 \caption{\label{fig:Fig_7}Excitonic gap parameter for intermediate (left panel) and large (right panel) values of the interlayer Coulomb interaction parameter $W/|t_{1}|$. The temperature is fixed at $T/|t_{1}|=0.5$.}
\end{center}
\end{figure}

In Figs.~\ref{fig:Fig_6} and ~\ref{fig:Fig_7} we have presented the excitonic gap parameter evolution for different values of the interlayer Coulomb interaction parameter $W/|t_{1}|$ and for different values of the interlayer electron hopping amplitude $t_{\perp}$. The temperature in all pictures in Figs.~\ref{fig:Fig_6} and ~\ref{fig:Fig_7} is fixed at the value $T/|t_{1}|=0.5$. We see in Figs.~\ref{fig:Fig_6} and ~\ref{fig:Fig_7} that for small values of the interlayer Coulomb interaction $W/|t_{1}|$, the excitonic gaps are very small. Following the evaluations from left to right in Figs.~\ref{fig:Fig_6} and ~\ref{fig:Fig_7}, we observe that the excitonic Frenkel-type gap formation peak is displacing to the region of the higher values of the intralayer Coulomb interaction parameter $U/|t_{1}|$ related to the fact that for the high values of $W/|t_{1}|$ the screening effects of $U/|t_{1}|$ are relatively small. The same is true for the merging point ``M'', where the two gap structures join together. It is clear in Figs.~\ref{fig:Fig_6} and ~\ref{fig:Fig_7} that the excitonic gap function also increases in the amplitude when increasing the interlayer coupling parameter $W/|t_{1}|$. This fact is related to the dominating role of the interlayer coupling between the particles. Meanwhile, the parameter $t_{\perp}$ is not affecting considerably the screening effects causing by the intralayer Coulomb interaction.    
%
%
%
%

\begin{figure}
\begin{center}
\includegraphics[scale=.6]{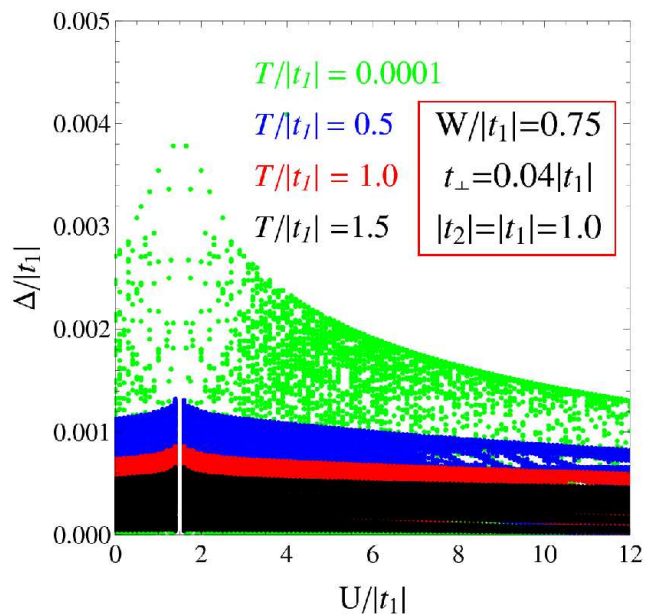}
\caption{\label{fig:Fig_8}(Color online) Excitonic gap parameter for the hopping-symmetric bilayer: $|t_{1}|=|t_{2}|=1$. Different values of the temperature are considered.}
\end{center}
\end{figure}

In Fig.~\ref{fig:Fig_8} we have shown the ecxitonic gap parameter for the case of the hopping-symmetric bilayer, when $|t_{1}|=|t_{2}|=1$. The solutions in  Fig.~\ref{fig:Fig_8} are obtained for the case $W/|t_{1}|=0.75$ and for the interlayer hopping amplitude $t_{\perp}=0.04|t_{1}|$. Different values of the temperature are considered. We see in Fig.~\ref{fig:Fig_8} that there is no Frenkel-type solution region and this is related to the high hopping value of the conduction band electrons in layer-2. Thus, by augmenting the layer-2 hopping amplitude, we suppress totally the regular Frenkel-type excitonic gap, and only the fluctuating Wannier-Mott-type region remains.   
%
%
\subsection{\label{sec:Section_4_2} Chemical potential}
\begin{figure}
\begin{center}
\includegraphics[scale=.5]{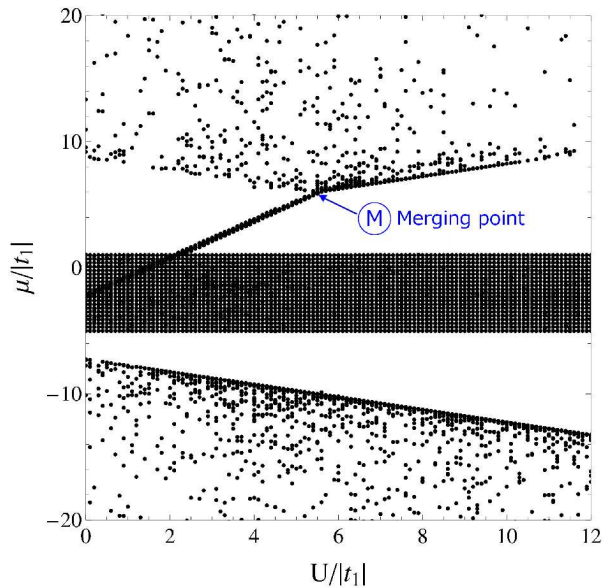}
\caption{\label{fig:Fig_9}(Color online) Chemical potential, at $T/|t_{1}|=0.0001$, normalized to the hopping amplitude $|t_{1}|$, as a function of the in-plane Coulomb interaction parameter $U/|t_{1}|$. The interlayer Coulomb interaction is set as $W/|t_{1}|=0.75$, and the interlayer hopping amplitude is $t_{\perp}=0.04|t_{1}|$.}
\end{center}
\end{figure}

%
In Fig.~\ref{fig:Fig_9} we have shown the solution for the chemical potential  for the case $W/|t_{1}|=0.75$, $t_{\perp}=0.04|t_{1}|$ and for $T/|t_{1}|=0.0001$. We see that the solution is given in the form of an asymmetric two-band structure, separated by the energy gap. Two bands of solutions have opposite sign, namely the upper band is positive and the lower band has the negative values. The upper band has the sense of the energies, at which the electron creation is possible in the layer-2, while the negative band is related to the energies of hole-doping in the system, in the layer-1. At the intermediate values of the intralayer Coulomb interaction $U/|t_{1}|$ the upper band solutions shoot down, to the region of the energy gap, and form a region located in the middle of the energy $\mu$-gap. By analyzing the solution of $\mu$ at $T/|t_{1}|=0.0001$, and corresponding values of the excitonic gap parameter $\Delta$ (see in Fig.~\ref{fig:Fig_4}), we conclude that the solution of $\mu$ in the middle of the $\mu$-gap is directly related to the degenerated Frenkel-type gap in the bilayer structure. 

\begin{figure}
\begin{center}
\includegraphics[scale=.53]{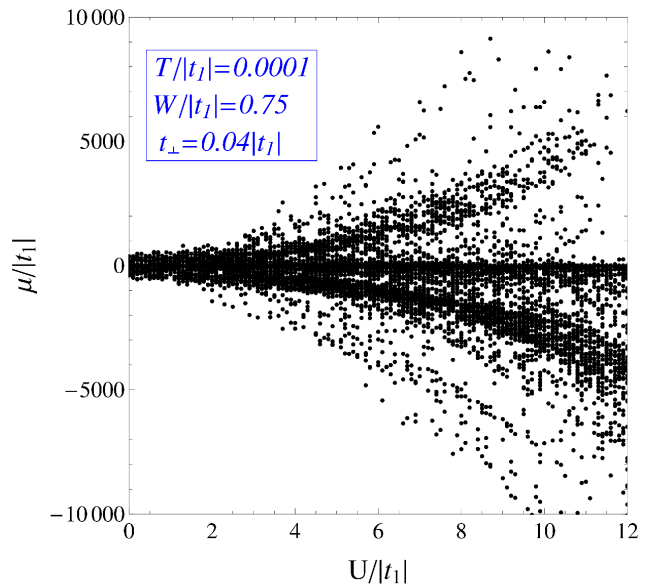}
\caption{\label{fig:Fig_10}(Color online) Chemical potential normalized to the hopping amplitude $|t_{1}|$, as a function of the in-plane Coulomb interaction parameter $U/|t_{1}|$. The interlayer Coulomb interaction is set as $W/|t_{1}|=0.75$, and the interlayer hopping amplitude is $t_{\perp}=0.04|t_{1}|$.}
\end{center}
\end{figure}

In Fig.~\ref{fig:Fig_10} we have presented the solutions for the chemical potential in the large energy scale, which straddle all solutions for the case of parameters $W/|t_{1}|=0.75$, $t_{\perp}=0.04|t_{1}|$ and $T/|t_{1}|=0.0001$. We see that the whole solution for the chemical potential forms a leaf-like structure. This is analogue to the similar structure for the chemical potential in the three-dimensional (3D) bulk semiconductors and which is discussed in Refs.\onlinecite{cite-29} and \onlinecite{cite-35}. Mainly, the chemical potential, forms a band, in 3D and 2D semiconducting materials, with the exciton formation possibility \cite{cite-35} and the upper bound of that solution is responsible for the excitonic gap parameter, and for the condensation of the preformed pairs.   
%

\begin{figure}
\begin{center}
\includegraphics[scale=.45]{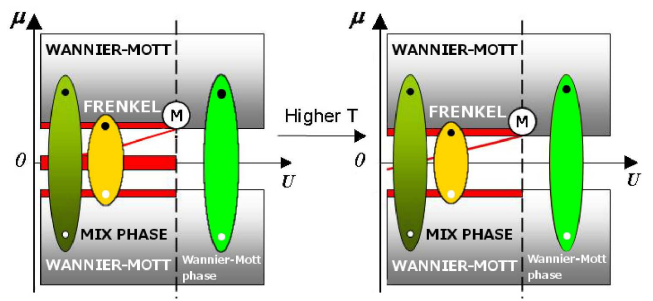}
\caption{\label{fig:Fig_11}(Color online) Schematic representation for the solution of the chemical potential in the quantum bilayer structure. The red color represents the solution of the chemical potential responsible for the formation of the Frenkel-type excitons. The left panel corresponds to the case of zero temperature and the right panel corresponds to higher temperatures.}
\end{center}
\end{figure}

%
Current discussion concerning the solution for the chemical potential  could be represented in the schematic form given in Fig.~\ref{fig:Fig_11}. We see in     
Fig.~\ref{fig:Fig_10} the regions, where, the Frenkel-type (the red region in the picture) or, the Wannier-Mott-type excitons (the region in gray) could be formed. The red regions in Fig.~\ref{fig:Fig_11}, correspond the values of chemical potential responsible for the formation of the Frenkel-type excitons. The red threads correspond to the upper solutions of the gap given in Figs.~\ref{fig:Fig_4} and ~\ref{fig:Fig_5}. 
%
%

\begin{figure}
\begin{center}
\includegraphics[scale=.53]{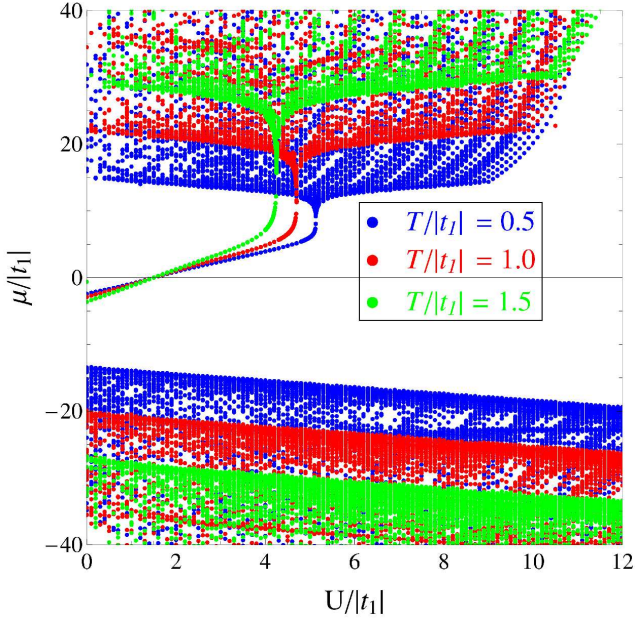}
\caption{\label{fig:Fig_12}(Color online) The temperature dependence of the chemical potential in the bilayer structure. The interlayer Coulomb interaction parameter is fixed at $W/|t_{1}|=0.75$ and $t_{\perp}=0.04|t_{1}|$. The beginning of the Frenkel-type solution (see the ``M''-point) is displaced to the smaller intralayer interaction region.}
\end{center}
\end{figure}

At the merging point ``M'' (see in Figs.~\ref{fig:Fig_9} and ~\ref{fig:Fig_11}) the Frenkel-type excitons are ``melting'' into the see of the degenerated Wannier-Mott low-energy region (see the low-energy regions in Figs.~\ref{fig:Fig_4} and ~\ref{fig:Fig_5}). For the strong intralayer Coulomb interactions $U/|t_{1}|$, the Frenkel-type excitations disappear, and the Wannier-Moot excitons are still present and this is corresponding to the vanishing of the Frenkel-type gap at the strong values of $U/|t_{1}|$ (see in Figs.~\ref{fig:Fig_4} and ~\ref{fig:Fig_5}).
The solution region at the middle of the $\mu$-gap, which is present in the case of very low temperatures (see in Fig.~\ref{fig:Fig_9} and also the left picture in Fig.~\ref{fig:Fig_11}) disappear, when augmenting the temperature (see the right picture in Fig.~\ref{fig:Fig_11} and also the solutions given in Fig.~\ref{fig:Fig_12}). 
%
%
%
\begin{figure}
\begin{center}
\includegraphics[scale=.53]{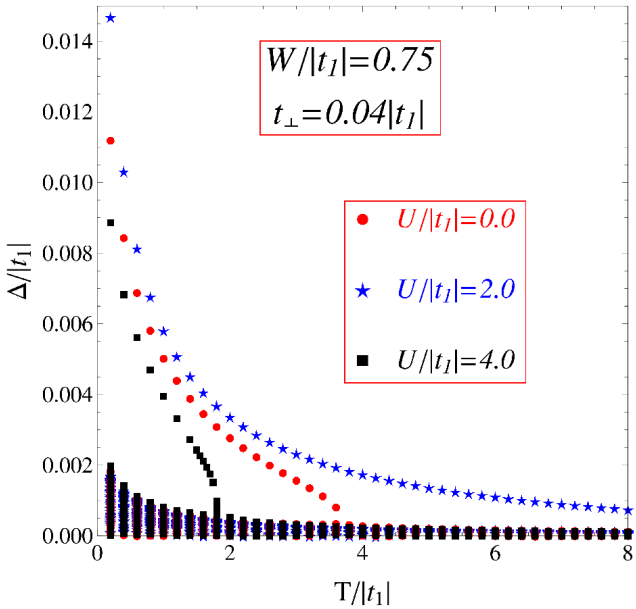}
\caption{\label{fig:Fig_13}(Color online) The temperature dependence of the excitonic gap parameter. Different values of the intralayer Coulomb interaction parameter $U/|t_{1}|$ are considered.}
\end{center}
\end{figure} 
%
\begin{figure}
\begin{center}
\includegraphics[width=200px, height=610px]{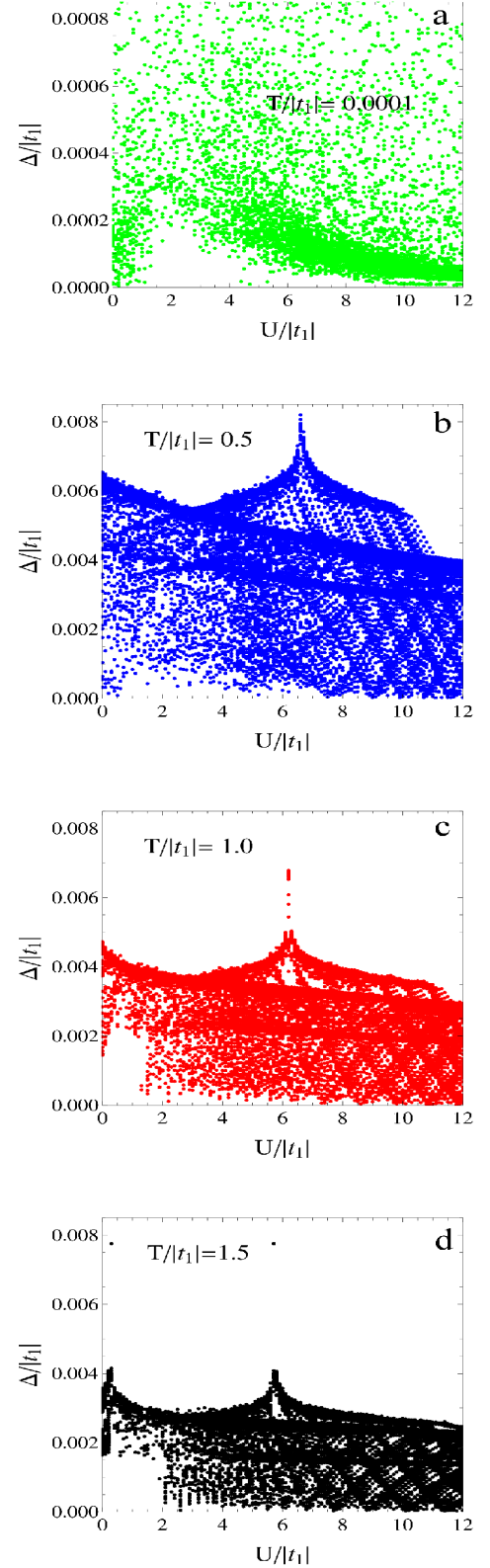}
\caption{\label{fig:Fig_14}(Color online)\scriptsize{ The enlarged Wannier-Mott-region of the bilayer excitonic structure, for different values temperature. The interlayer Coulomb interaction parameter is fixed at $W/|t_{1}|=1.5$ and the interlayer hopping amplitude is set as $t_{\perp}=0.1|t_{1}|$. The figures in panels a-d, show the very small Wannier-Mott-type gap-lobs, which are opening in the system.}}
\end{center}
\end{figure} 
%
In Fig.~\ref{fig:Fig_12} we give the temperature dependence of the chemical potential for the case $W/|t_{1}|=0.75$ and $t_{\perp}=0.04|t_{1}|$. Different colors in the picture correspond to different values of renormalized temperature $T/|t_{1}|$. The blue points represent the solutions for $\mu$ at $T/|t_{1}|=0.5$, red points - for $T/|t_{1}|=1.0$, and green points - for $T/|t_{1}|=1.5$. It is clear that the temperature kills the solution region in the middle of the $\mu$-gap, which is present in the case of very low temperatures $T/|t_{1}|=0.0001$. We see in In Fig.~\ref{fig:Fig_12}, that with increasing the temperature the $\mu$-band separation increases and the Frenkel-type solution with the ``M''-point is displaced to the region of small intralayer interaction region, in accordance with the similar behavior of the merging point ``M'' in the excitonic gap structure in Figs.~\ref{fig:Fig_4} and ~\ref{fig:Fig_5}. 

In Fig.~\ref{fig:Fig_13} we have presented the temperature dependence of the excitonic gap  parameter (both, the Frenkel-type, and Wannier-Moot-type solutions are given) for the case $W/|t_{1}|=0.75$ and small interlayer hopping parameter $t_{\perp}=0.04|t_{1}|$. We see in Fig.~\ref{fig:Fig_13} that, when increasing the intralayer Coulomb interaction parameter, the gap function also increases, then for the medium values of the Coulomb interaction $U/|t_{1}|=4$, it decreases and finally disappears at the large values of $U/|t_{1}|$. This behavior is consistent with the behavior of the excitonic gap function given in Figs.~\ref{fig:Fig_4} and ~\ref{fig:Fig_5}.  

Furthermore, in the Fig.~\ref{fig:Fig_14}, we show how very small Wannier-Moot gap lobs are opening in the spectrum of the excitonic gap function. For this, we consider the lowest gap structure region (see the lowest parts in Figs.~\ref{fig:Fig_4} and ~\ref{fig:Fig_5}). On the large-scale pictures in Fig.~\ref{fig:Fig_14} we see how the very small Wannier-Mott-type gap islands are opening in the gap structure, when increasing the intralayer Colomb repulsion. The amplitudes of these small gaps are decreasing, when increasing $U/|t_{1}|$ and vanish, when the interlayer Coulomb screening is very high.  For the very low temperatures (see in the panel-a in  Fig.~\ref{fig:Fig_14}) we have only one well defined Mott-gap slope, which is spliced in many lobes, when increasing the temperature (see the panels-b, -c and -d in Fig.~\ref{fig:Fig_14}).
%

\section{\label{sec:Section_5} Final remarks and conclusions}
%
We have considered the electronic double layer structure without the applied external electric field. Employing the effective action formalism we have shown the complicated excitonic gap structure in the bilayer system. We demonstrated that both Frenkel and Wannier-Mott-type gaps are possible in the strongly correlated bilayer system. Considering the chemical potential structure, we are able to separate all possible gap formation regions and discuss the related energy scales. For the Frenkel-type gap, we have found a very narrow region in the chemical potential structure, where the Frenkel-type excitons could form. 

Meanwhile, we have shown that the Frenkel-type excitons correspond to the very small values of the chemical potential, contrary to the case of the Wannier-Mott low excitation region, where the formed excitons correspond to the very large particle $\mu$-energy scales.

The results show that the single-particle energies at which the Wannier-Mott-type excitons could be formed are relatively very high, and, energetically, it appears that the Frenkel-type solution is more favored for the bilayer structure. The calculated $U$-dependence of the excitonic gap function shows that for the small and medium values of the Coulomb interaction parameter $U/|t_{1}|$ both, Frenkel and Wannier-Mott-type exciton formations are possible, while in the strong intralayer interaction case the Frenkel-gap is vanishing rapidly, while the Wannier-Mott gap survives at relatively high values of the Coulomb interaction parameter $U/|t_{1}|$. For the future, it would be interesting to consider the double layered structure in the case of the non-zero external electric field and to see the modifications that occur in the gap structure for that case. We suppose that the non-zero applied electric field will favor the exciton formation processes in the bilayer system and we expect that the amplitudes of the excitonic gap parameters should be much higher in that case. Furthermore, we are planning to discuss the possibility of the exciton condensation  in the bilayer type systems.  
%



\begin{thebibliography}*
%
\bibitem{cite-1} E. McCann, Phys.Rev.B \textbf{74}, 161403 (2006).
\bibitem{cite-2} M. Mucha-Kruczynski, E. McCann, and V.I. Fal'ko, Semicond.Sci.Technol. \textbf{25}, 033001 (2010).
\bibitem{cite-3} J. Nilsson, A.H. Castro Neto, F. Guinea amd N.M.R. Peres, Phys.Rev.B \textbf{78}, 045405 (2008).
\bibitem{cite-4} P.Gava, M. Lazzeri, A.M. Saitta, and F.Mauri, Phys.Rev.B \textbf{79}, 165431 (2009).
\bibitem{cite-5} T.Ohta, A.Bostwick, T. Seyller, K.Horn, and E. Rotenberg, Science \textbf{313}, 951 (2006).
\bibitem{cite-6} L. Rademaker, K.Wu, H. Hilgenkamp, and J. Zaanen, Europhys. Lett. \textbf{97}, 27004 (20012).
\bibitem{cite-7} H.Min, R. Bistritzer, J.J.Su, and A.H. MacDonald, Phys.Rev.B \textbf{78}, 121401(R) (2008).
\bibitem{cite-8} N.V. Phan and H. Fehske, New J. Phys. \textbf{14}, 075007 (2012).
\bibitem{cite-9} A. Perali, D.Neilson, and A.R. Hamilton, Phys.Rev.Lett. \textbf{110}, 146803 (2013).
\bibitem{cite-10} K.Zou, X.Hong, and J.Zhu, Phys.Rev.B, 84, 085408 (2011).
\bibitem{cite-11} K. Wu, L. Rademaker, and J. Zaanen, Phys. Rev. Appl. \textbf{2}, 054013 (2014).
\bibitem{cite-12} Y. Naveh and B. Laikhtman, Phys.Rev.Lett. \textbf{77}, 900 (1996).
\bibitem{cite-13} X. Zhu, P.B.Littlewood, M.S.Hybertsen, and T.M.Rice, Phys.Rev.Lett. \textbf{74}, 1633 (1994).
\bibitem{cite-14} B.Seradjeh, J.E.Moore, and M.Franz, Phys.Rev.Lett. \textbf{103}, 066402 (2009).
\bibitem{cite-15}J.P. Eisenstein and A.H.MacDonald, Nature, \textbf{432}, 691 (2004).
\bibitem{cite-16} T. Kaneko, S.Ejima, H.Fehske, and Y.Ohta, Phys.Rev.B, 
\bibitem{cite-17} D.I. Pikulin and T. Hyart, Phys.Rev.Lett. \textbf{112}, 176403 (2014).
\bibitem{cite-18} L. Rademaker, J. van den Brink, J. Zaanen, and H. Hilgenkamp, Phys.Rev.B \textbf{88}, 235127 (2013). 
\bibitem{cite-19} L. Rademaker, J., Zaanen, and H. Hilgen, Phys.Rev.B \textbf{83}, 012504 (2011). 
\bibitem{cite-20} L. V. Keldysh and Y. V. Kopaev, Fiz. Tverd. Tela Leningrad \textbf{6}, 2791 (1964) [Sov. Phys. Solid State \textbf{6}, 2219 (1965)].
\bibitem{cite-21} S. A. Moskalenko, Fiz. Tverd. Tela Leningrad \textbf{4}, 276 (1962) [Sov. Phys. Solid State \textbf{4}, 199 (1962)].
\bibitem{cite-22} P.B.Littlewood, P.R. Eastham, J.M.J. Keeling, F.M. Marchetti, B.D.Simons, and M.H.Szymanska, J. Phys. Condens. Matter 16, S3597 (2004).
\bibitem{cite-23} J.M.Blatt, K.W. Boer, and W. Brandt, Phys.Rev. \textbf{126}, 1691 (1962).
\bibitem{cite-24} D. Snoke, Science \textbf{15}, 1368, (2002).
 S. A. Moskalenko and D. W. Snoke, \textit{Bose-Einstein Condensation
of Excitons and Biexcitons} (Cambridge Univ. Press, Cambridge, 2000).
\bibitem{cite-25}  D. Ihle, M. Pfafferott, E. Burovski, F. X. Bronold, and H. Fehske, Phys. Rev. B \textbf{78}, 193103 (2008).
\bibitem{cite-26} B. Zenker, D. Ihle, F. X. Bronold, and H. Fehske, Phys.
Rev. B \textbf{81}, 115122 March (2010).
\bibitem{cite-27} B. Zenker, D. Ihle, F. X. Bronold, and H. Fehske, Phys. Rev. B \textbf{83}, 235123 (2011).
\bibitem{cite-28}B. Zenker, D. Ihle, F. X. Bronold, and H. Fehske, Phys. Rev. B \textbf{85}, 121102 (2012).
\bibitem{cite-29}  K. Seki, R. Eder, and Y. Ohta, Phys. Rev. B, \textbf{84}, 245106 (2011).
\bibitem{cite-30} Yuh Tomio, Kotaro Honda, and Tetsuo Ogawa, Phys. Rev. B \textbf{73}, 235108, (2006) .
\bibitem{cite-31} D.W. Snoke, Coherence and Optical Emission from Bilayer Exciton Condensates, Advances in Condensed Matter Physics 2011, 938609 (2011).
\bibitem{cite-32} T. Fukuzawa, E.E.Mendez, and J.M. Hong, Phys.Rev.Lett. \textbf{64}, 3066 (1990).
\bibitem{cite-33} J.P.-Cheng, J.Kono, B.D.McCombe, I. Lo, W.C. Mitchel. and C.E.Stutz, Phys.Rev.Lett. \textbf{74}, 450 (1995).
\bibitem{cite-34} J. Szyma\'{n}ski, L.\'{S}wierkowski and D. Neilson, Phys.Rev.B \textbf{50}, 11002 (1994). 
 \bibitem{cite-35} V. Apinyan and T.K. Kope\'{c}, Journal of Low Temperature Physics, \textbf{176}, 27 (2014).
\bibitem{cite-36} V. Apinyan and T.K. Kope\'{c}, Journal of Low Temperature Physics, \textbf{178}, 295 (2015).
\bibitem{cite-37} J. W. Negele and H. Orland, \textit{Quantum Many-Particle Systems}, Addison-Wesley, Reading, MA, (1988).
\bibitem{cite-38} A.A. Abrikosov, L.P. Gorkov, I.E. Dzyaloshinski, \textit{Methods of Quantum Field Theory in Statistical Physics}, Pergamon Press (1965).
\bibitem{cite-39} M.J.D. Powell, A hybrid method for nonlinear equations, in numerical methods of nonlinear algebraic equations, Rabinowitz, P. ed., Gordon and Breach, New York (1970).

\end{thebibliography}
\end{document}